\documentstyle[twocolumn,seceq,graphics,graphicx]{jpsj}

\def\runtitle{Instructions for the Preparation of Manuscript}
\def\runauthor{Tomohiko {\sc Nakajima}, Hideki {\sc Yoshizawa} and Yutaka {\sc Ueda}}

\title
{
$A$-site Randomness Effect on Structural and Physical Properties of Ba-based Perovskite Manganites
}

\author
{ 
Tomohiko {\sc Nakajima}$^{1}$\footnote{Present address: Materials Design and Characterization Laboratory, Institute for Solid State Physics, the University of Tokyo, 5-1-5 Kashiwanoha, Kashiwa, Chiba 277-8581, Japan}, Hideki {\sc Yoshizawa}$^{2}$ and Yutaka {\sc Ueda}$^{1}$
}

\inst
{
$^1$Materials Design and Characterization Laboratory, Institute for Solid State Physics, the University of Tokyo, 5-1-5 Kashiwanoha, Kashiwa, Chiba 277-8581, Japan\\
$^2$Neutron Science Laboratory, Institute for Solid State Physics, University of Tokyo, 106-1 Shirakata, Tokai, 319-1106, Japan\\
}

\recdate
{
\today
}

\abst
{
The discovery of novel structural and physical properties in the $A$-site 
ordered manganite $R$BaMn$_{2}$O$_{6}$ ($R$ = Y and rare earth elements) has 
demanded new comprehension about perovskite manganese oxides. In the present 
study, the $A$-site disordered form, $R_{0.5}$Ba$_{0.5}$MnO$_{3}$, has been 
investigated and compared with both $R$BaMn$_{2}$O$_{6}$ and 
$R_{0.5}A_{0.5}$MnO$_{3}$ ($A$: Sr, Ca) in the structures and electromagnetic 
properties. $R_{0.5}$Ba$_{0.5}$MnO$_{3}$ has a primitive cubic perovskite 
cell in the structure and magnetic glassy states are dominant as its ground 
state, in contrast to the ordinary disordered 
$R_{0.5}A_{0.5}$MnO$_{3}$ ($A$: Sr, Ca). In Pr-compounds with various degrees 
of Pr/Ba randomness at the $A$-sites, the $A$-site disorder gradually suppresses 
both ferromagnetic and A-type antiferromagnetic transitions and finally 
leads to a magnetic glassy state in Pr$_{0.5}$Ba$_{0.5}$MnO$_{3}$. A 
peculiar behavior, multi-step magnetization and resistivity change, has been 
observed in Pr$_{0.5}$Ba$_{0.5}$MnO$_{3}$. These properties could be closely 
related to any spatial heterogeneity caused by the random distribution of 
Ba$^{2 + }$ and $R^{3 + }$ with much different ionic radius.
}

\kword
{
A-site ordered/disordered perovskite manganites, Crystal structure, Electromagnetic properties, Randomness effect, Magnetoresistance effect
}

\begin{document}
\sloppy
\maketitle

\section{Introduction}
\begin{figure}[t!]
\begin{center}
\includegraphics{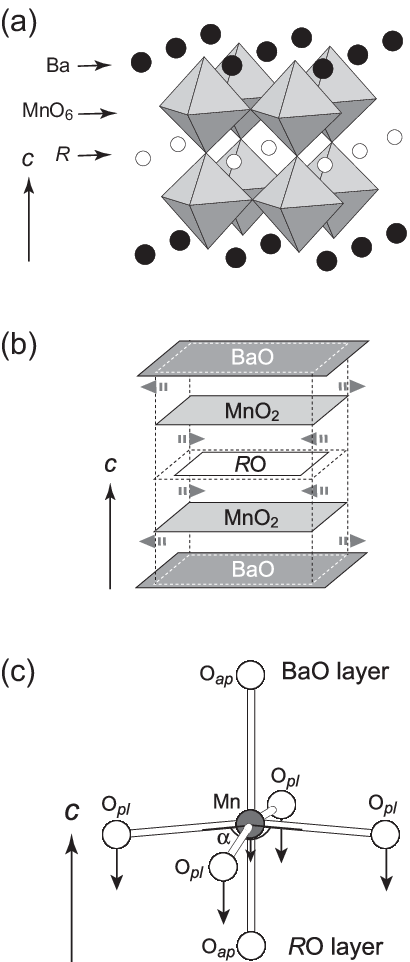}
\end{center}
\caption{(a) Crystal structure and (b) structural concept of the $A$-site ordered 
manganite $R$BaMn$_{2}$O$_{6}$, and (c) an illustration of the distorted 
MnO$_{6}$ octahedron.}
\label{f1}
\end{figure}

\begin{figure}[t!]
\begin{center}
\includegraphics{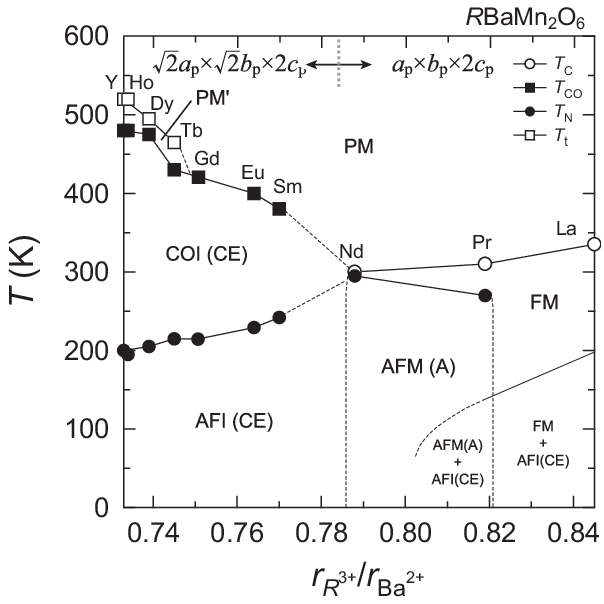}
\end{center}
\caption{Electronic phase diagram for the $A$-site ordered manganite 
$R$BaMn$_{2}$O$_{6}$. PM (PM'): paramagnetic metal phase, FM: ferromagnetic 
metal phase, AFM(A): A-type antiferromagnetic metal phase, COI(CE): CE-type 
charger/orbital ordered insulator phase, AFI(CE): CE-type antiferromagnetic 
insulator phase.}
\label{f2}
\end{figure}

The magnetic and electrical properties of perovskite manganites with the 
general formula ($R^{3 + }_{1 - x}A^{2 + }_{x})$MnO$_{3}$ ($R$ = rare earth 
elements and $A$ = Ca and Sr) have been extensively investigated for the last 
decade.$^{1)}$ Among the interesting features are the so-called colossal 
magnetoresistance (CMR) and metal-insulator (MI) transition accompanied by 
charge/orbital order (CO). It is now widely accepted that these enchanting 
phenomena are caused by the strong correlation/competition among spin, 
charge and orbital degrees of freedom, which would be significantly 
influenced by the $A$-site randomness. Recently we successfully synthesized the 
$A$-site ordered manganite, $R$BaMn$_{2}$O$_{6}$ ($R$ = Y and rare earth elements) 
and reported its structure and electromagnetic properties.$^{2 - 6)}$ As 
schematically shown in Fig. \ref{f1}, the most significant structural feature of 
$R$BaMn$_{2}$O$_{6}$ is that the MnO$_{2}$ square sublattice is sandwiched by 
two types of rock-salt layers, $R$O and BaO, with much different sizes, and 
consequently the MnO$_{6}$ octahedron itself is distorted in a 
noncentrosymmetric manner that both Mn and oxygen atoms in the MnO$_{2}$ 
plane are displaced toward the $R$O layer (Fig. \ref{f1}(c)), in contrast to the rigid 
MnO$_{6}$ octahedron in the $A$-site disordered manganite ($R_{1 - 
x}A_{x})$MnO$_{3}$ ($A$ = Ca and Sr). This means that the structural and 
physical properties of $R$BaMn$_{2}$O$_{6}$ can be no longer explained in terms 
of the basic structural distortion, the so-called tolerance factor $f$, as can 
be in ($R_{0.5}A_{0.5})$MnO$_{3}$ ($A$ = Ca and Sr).$^{1)}$ Figure \ref{f2} shows the 
electronic phase diagram of $R$BaMn$_{2}$O$_{6}$ expressed as a function of the 
ratio of ionic radius of the $A$-site cations.$^{3)}$ Among possible 
combinations of $R$/Ba, the mismatch between $R$O and BaO is the smallest in La/Ba 
and the largest in Y/Ba. As seen in Fig. \ref{f2}, the CE-type charge/orbital 
ordered state (COI(CE)) with a new stacking variation of the CE-type CO is 
stabilized at the relatively high temperatures ($T_{\rm CO})$ far above 300 K, 
when $R^{3 + }$ is smaller than Sm$^{3 + }$ in ion size. The high $T_{\rm CO}$ 
would be not only due to the absence of $A$-site randomness but also due to the 
distorted structure with a tilt of MnO$_{6}$ octahedra as well as heavy 
distortion of MnO$_{6}$ octahedron.$^{5)}$ The new CE-type CO with a 4-fold 
periodicity along the $c$-axis (4CE-CO) could be due to a layer type order of 
$R$ and Ba.$^{4)}$ Interestingly, this 4CE-CO changes into a new type with a 
two or single periodicity along the $c$-axis, when the system enters into the 
antiferromagnetic CE-type charge/orbital ordered state (AFI(CE)).$^{4)}$ 
Furthermore, $R$BaMn$_{2}$O$_{6}$ ($R$ = Tb, Dy, Ho and Y) shows the structural 
transition at $T_{\rm t}$ above $T_{\rm CO}$, as shown in Fig. \ref{f2}, which is possibly 
accompanied by a $d_{x^2 - y^2} $ type orbital order.$^{3,5)}$ Therefore the 
degeneracy of orbital, charge and spin degrees of freedom are lifted in 
these compounds. On the other hand, $R$BaMn$_{2}$O$_{6}$ ($R$ = La, Pr and Nd) 
with relatively larger $R^{3 + }$ has no octahedral tilt and shows a 
transition from a paramagnetic metal (PM) to a ferromagnetic metal (FM). The 
ground states for PrBaMn$_{2}$O$_{6}$ and NdBaMn$_{2}$O$_{6}$ are the A-type 
antiferromagnetic metal (AFM(A)). In LaBaMn$_{2}$O$_{6}$, AFI(CE) phase 
coexists with FM phase in the ground state, which suggests that the 
electronic phase separation is not due to the $A$-site randomness but is 
intrinsic phenomenon in perovskite manganites where FM and CO interactions 
compete against each other and are significantly influenced by a tiny change 
of the local structure.$^{6)}$

The discovery of such novel structural and physical properties in the 
$A$-site ordered manganite $R$BaMn$_{2}$O$_{6}$ has demanded new comprehension 
about perovskite manganese oxides.$^{3 - 11)}$ Recently, theoretical studies 
also have revealed that the interesting properties such as CMR and 
electronic phase separation come from a critical competition between FM and 
an antiferromagnetic CO interaction, which could be significantly influenced 
by the $A$-site randomness or a fluctuation of the local structure$^{12,13)}$ 
The $A$-site disordered form ($R_{0.5}$Ba$_{0.5})$MnO$_{3}$ with the same 
constituent elements is crucial to deepen the understanding of the 
structural and physical properties of perovskite manganites; it may make 
clear the effects of $A$-site randomness not only qualitatively but also 
quantitatively. Very recently Akahoshi \textit{et al.} reported that the magnetic glassy 
state became dominant in ($R_{0.5}$Ba$_{0.5})$MnO$_{3}$.$^{11)}$ However the 
detailed structure and electromagnetic properties of 
($R_{0.5}$Ba$_{0.5})$MnO$_{3}$ has not been reported. We have also 
independently synthesized the $A$-site disordered Ba-based manganite 
($R_{0.5}$Ba$_{0.5})$MnO$_{3}$ and studied the structure and electromagnetic 
properties in terms of the degree of $A$-site randomness. In this paper, we 
report the structures and electromagnetic properties of 
($R_{0.5}$Ba$_{0.5})$MnO$_{3}$, especially Pr-compounds with various degrees 
of the randomness of Pr/Ba at the $A$-sites, and we will discuss the obtained 
results in terms of the $A$-site randomness effect.

\section{Experimental}
\begin{figure}
\begin{center}
\includegraphics{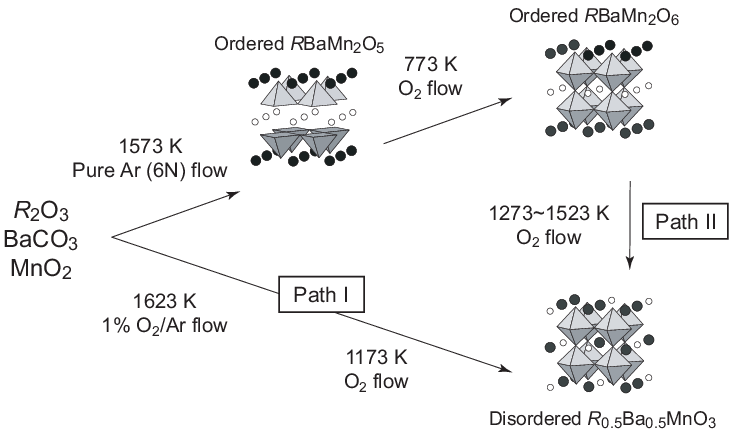}
\end{center}
\caption{Flowchart of sample preparation.}
\label{f3}
\end{figure}

Polycrystalline samples of ($R_{0.5}$Ba$_{0.5})$MnO$_{3}$ ($R$ = Y and rare 
earth elements) were prepared by a solid-state reaction of $R_{2}$O$_{3}$, 
BaCO$_{3}$ and MnO$_{2}$ at 1623 K in 1{\%} O$_{2}$/Ar gas for 1 day, 
followed by annealing at 1173 K in O$_{2}$ gas for 1 day (Path I in Fig. \ref{f3}). 
The preparation method of the ordered form $R$BaMn$_{2}$O$_{6}$ was reported 
elsewhere.$^{2,3,5)}$ Interestingly, annealing $R$BaMn$_{2}$O$_{6}$ under 
O$_{2}$ gas at high temperatures always resulted in insufficient $R$/Ba 
solid-solution at the $A$-sites. Pr-compounds: [Pr$_{g}$Ba$_{1 - 
g}$]$_{\rm P}$[Pr$_{1 - g }$Ba$_{g}$]$_{\rm B}$Mn$_{2}$O$_{6}$ (0.5$ \le g \le 
$1.0) with various degrees of the $A$-site order were synthesized from 
PrBaMn$_{2}$O$_{6}$ by controlling the annealing time and temperatures 
(1273$\sim $1623 K) in O$_{2}$ gas (Path II in Fig. \ref{f3}), where [ ]$_{\rm P}$ (or 
[ ]$_{\rm B})$ represents Pr-sites (or Ba-sites) in PrBaMn$_{2}$O$_{6}$. The 
degree of $A$-site order $(S)$, $S$ = (2g-1) $\times $ 100 {\%}, was determined by 
the Rietveld analysis of powder X-ray and neutron diffractions. We obtained 
$S$ = 96$\pm $2{\%} for the ordered form PrBaMn$_{2}$O$_{6}$ and $S$ = 0.0 {\%} 
for the disordered form Pr$_{0.5}$Ba$_{0.5}$MnO$_{3}$ prepared by Path I in 
Fig. \ref{f3}. In the present study, we successfully prepared Pr-compounds: 
[Pr$_{g}$Ba$_{1 - g}$]$_{\rm P}$[Pr$_{1 - g }$Ba$_{g}$]$_{\rm B}$Mn$_{2}$O$_{6}$ 
(0.5$ \le g \le $1.0) with $S$ = 96(2){\%}, 87(4){\%}, 70(4){\%}, 57(6){\%}, 
32(2){\%}, 25(6){\%} and 0.0 {\%} which were named as PB96, PB87, PB70, 
PB57, PB32, PB25 and PB00, respectively. The synthesis conditions of these 
Pr-compounds are shown in Table \ref{t1}.

\begin{fulltable}[htbp]
\caption{Synthesis conditions of Pr-compounds (PB96$\sim$PB00) with various degrees of the $A$-site randomness. (See the text).}
\label{t1}
\begin{fulltabular}{@{\hspace{\tabcolsep}\extracolsep{\fill}}cl} \hline
Sample&Synthesis condition\\
\hline
PB96 & Heating starting materials at 1623 K in pure Ar for 24 h, followed by annealing at 623 K in O$_2$ for 24 h\\\\
PB87 & Annealing PB96 at 1273 K in O$_2$ for 6 h\\\\
PB70 & Annealing PB87 at 1523 K in O$_2$ for 24 h\\\\
PB57 & Annealing PB70 at 1523 K in O$_2$ for 24 h\\\\
PB32 & Annealing PB57 at 1623 K in O$_2$ for 24 h\\\\
PB25 & Annealing PB32 at 1623 K in O$_2$ for 24 h\\\\
PB00 & Heating starting materials at 1623 K in 1\% O$_2$/Ar for 24 h, followed by annealing at 1173 K in O$_2$ for 24 h\\
\hline
\\
\end{fulltabular}
\end{fulltable}

The crystal structures including the order/disorder of $R$/Ba were refined by 
the Rietveld analysis of powder X-ray and neutron diffractions using RIETAN 
2000$^{14)}$. The magnetic ordered states at low temperatures were studied 
by powder neutron diffraction. The X-ray powder diffraction experiments were 
performed using a MXP21 Mac Science diffractometer with the following 
operation conditions: 5\r{}$<$ 2\textit{$\theta $} $<$120\r{} with the step size of 0.02\r{ }, 
Cu-\textit{K$\alpha $} radiation, $V$ = 45 kV and $I$ = 350 mA. The powder neutron diffraction was 
performed with the Kinken powder diffractometer, HERMES, of Institute for 
Materials Research (IMR), Tohoku University, installed at the JRR-3M reactor 
in Japan Atomic Energy Research Institute (JAERI), Tokai. Neutrons with a 
wavelength of 1.8207 {\AA} were obtained by the 331 reflection of the Ge 
monochromator and the 12'-blank-sample-22' collimation. The magnetic 
properties were studied using a SQUID magnetometer in a temperature range 
$T$ = 2$\sim $400 K. The electric resistivity of a sintered pellet was measured 
for $T$ = 2$\sim $400 K by a conventional four-probe technique.

\section{Results and discussion}
\subsection{The $A$-site disordered $R_{0.5}$Ba$_{0.5}$MnO$_{3}$}

\begin{figure}
\begin{center}
\includegraphics{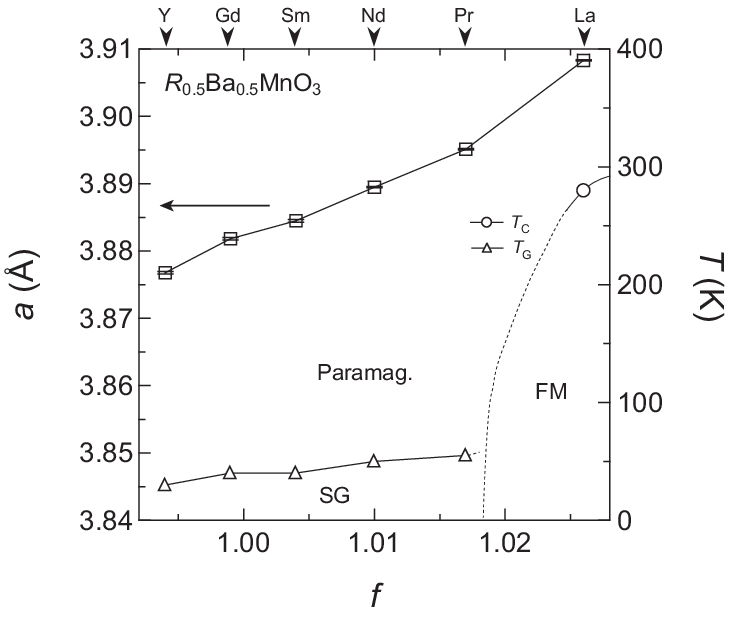}
\end{center}
\caption{Lattice parameters at room temperature and phase diagram of the $A$-site 
disordered $R_{0.5}$Ba$_{0.5}$MnO$_{3}$ as a function of the tolerance factor 
$f$. SG: spin glass phase.}
\label{f4}
\end{figure}

\begin{figure}
\begin{center}
\includegraphics{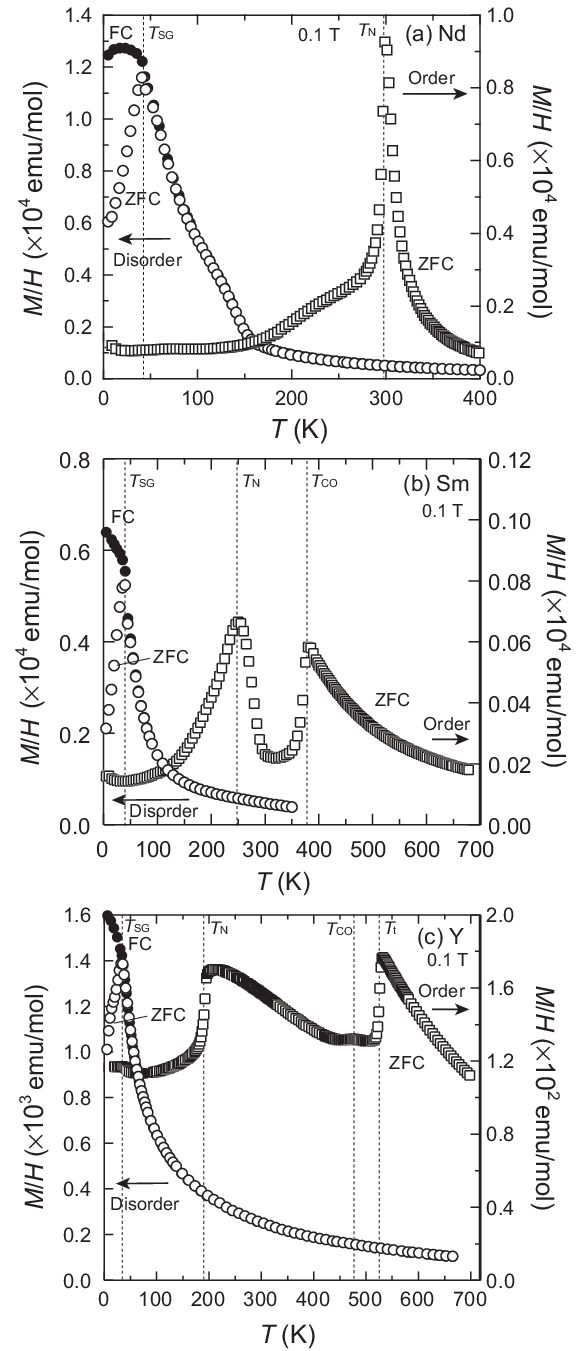}
\end{center}
\caption{Temperature dependence of magnetic susceptibility for the $A$-site 
ordered/disordered (a) NdBaMn$_{2}$O$_{6}$/Nd$_{0.5}$Ba$_{0.5}$Mn-O$_{3}$, 
(b) SmBaMn$_{2}$O$_{6}$/Sm$_{0.5}$Ba$_{0.5}$MnO$_{3}$ and (c) 
YBaMn$_{2}$O$_{6}$/-Y$_{0.5}$Ba$_{0.5}$MnO$_{3}$ under 0.1 T.}
\label{f5}
\end{figure}

\begin{figure}
\begin{center}
\includegraphics{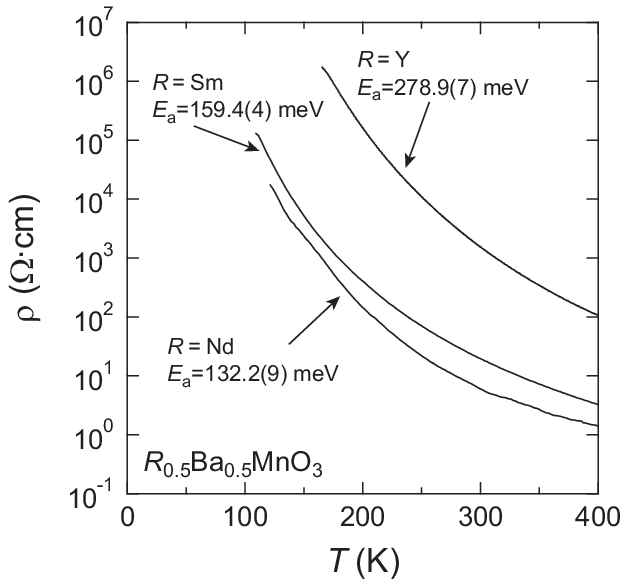}
\end{center}
\caption{Temperature dependence of resistivity for the sintered 
$R_{0.5}$Ba$_{0.5}$MnO$_{3}$ ($R$ = Nd, Sm and Y) sample.}
\label{f6}
\end{figure}

The X-ray diffraction patterns of all $R_{0.5}$Ba$_{0.5}$MnO$_{3}$ can be 
indexed in the primitive cubic perovskite cell. There is no extra peak 
suggesting a superstructure. The lattice parameters of 
$R_{0.5}$Ba$_{0.5}$MnO$_{3}$ at room temperature are shown in Fig. \ref{f4}. The 
lattice parameter decreases with decreasing ionic radius of $R^{3 + }$. The 
simple cubic cell means no tilt of MnO$_{6}$ octahedra in contrast to the 
orthorhombic GdFeO$_{3}$ type distortion of $R_{0.5}A_{0.5}$MnO$_{3}$ ($A$ = Ca 
and Sr). In general, the mismatch between the larger MnO$_{2}$ and the 
smaller ($R$,$A)$O sublattices is relaxed by tilting MnO$_{6}$ octahedra, 
resulting in the lattice distortion from cubic to, mostly, the orthorhombic 
GdFeO$_{3}$-type structure. In this distortion, the bond angle $\angle 
$Mn-O-Mn deviates from 180\r{ }, leading to a significant change in the 
effective one-electron bandwidth or equivalent $e_{g}$-electron transfer 
interaction, and the degree of this mismatch is described as 
$f = (\langle r_{A} \rangle + r_{\rm O} )/\sqrt2 [(r_{\rm Mn} + r_{\rm O})]$
, where $\langle r_{A} \rangle$, $r_{\rm Mn}$ and $r_{\rm O}$ are 
(averaged) ionic radii for the respective elements. The electronic phase 
diagram of $R_{0.5}A_{0.5}$MnO$_{3}$ ($A$ = Ca and Sr) has been explained by 
$f$; FM state generated by the double-exchange interaction is dominant near $f$ = 
1, while COI(CE) state is most stabilized in the lower $f$ region ($f $$<$ 
0.975).$^{1)}$ In $R_{0.5}$Ba$_{0.5}$MnO$_{3}$, the $f$ is in the range from 
1.026 (La/Ba) to 0.995 (Y/Ba), which are rather close to $f$ = 1, comparing to 
the variation $0.955 < f < 1$ in $R_{0.5}A_{0.5}$MnO$_{3}$ ($A$ = Ca and 
Sr).$^{15)}$ Therefore the simple cubic structures of 
$R_{0.5}$Ba$_{0.5}$MnO$_{3}$ can be partly understood from the $f$-values close 
to 1, that is relatively small mismatch between MnO$_{2}$ and 
($R_{0.5}$Ba$_{0.5})$O lattices. Here, it should be noticed that the $f$ is 
beyond 1 in $R_{0.5}$Ba$_{0.5}$MnO$_{3}$ ($R$ = La, Pr and Nd). Actually the 
lattice parameters (3.904 $ \sim $ 3.918 {\AA})$^{6)}$ of these compounds 
are larger than the ideal one ($ \sim $3.89 {\AA})$^{15)}$ of Mn$^{3.5 + 
}$O$_{2}$ lattice. From simple cubic structures of 
$R_{0.5}$Ba$_{0.5}$MnO$_{3}$, one may expect FM generated by double exchange 
interaction as the stable electronic state. The ground state of 
La$_{0.5}$Ba$_{0.5}$MnO$_{3}$ is actually a pure FM and the ferromagnetic 
transition temperature $T_{\rm C}$ decreases by 50 K compared with $T_{\rm C}$ = 330 K 
in LaBaMn$_{2}$O$_{6}$, agreeing with the previous report$^{7)}$. On the 
other hand, Pr$_{0.5}$Ba$_{0.5}$MnO$_{3}$ and Nd$_{0.5}$Ba$_{0.5}$MnO$_{3}$ 
show the increase of magnetic susceptibility ($M/H)$ below about 150 K and then 
show glassy behaviors below about 50 K, evidenced by significant differences 
of $M/H$-$T$ curves on zero-field cooled (ZFC) and field cooled (FC) processes. As 
an example, the $M/H$-$T$ curve for Nd$_{0.5}$Ba$_{0.5}$MnO$_{3}$ is shown in Fig. 
\ref{f5}(a), together with that for NdBaMn$_{2}$O$_{6}$. Akahoshi \textit{et al.}$^{11)}$ 
previously reported FM states for Pr$_{0.5}$Ba$_{0.5}$MnO$_{3}$ and 
Nd$_{0.5}$Ba$_{0.5}$MnO$_{3}$, which could be due to imperfect disorder as 
verified in the following section of this paper. More typical spin-glass 
behaviors have been observed in $R_{0.5}$Ba$_{0.5}$MnO$_{3}$ with Sm$^{3 + }$ 
and smaller $R^{3 + }$s, agreeing with the previous report$^{11)}$. The 
typical $M/H$-$T$ curves of Sm$_{0.5}$Ba$_{0.5}$MnO$_{3}$ and 
Y$_{0.5}$Ba$_{0.5}$MnO$_{3}$ are shown in Fig. \ref{f5}(b) and \ref{f5}(c) together with 
those for SmBaMn$_{2}$O$_{6}$ and YBaMn$_{2}$O$_{6}$. YBaMn$_{2}$O$_{6}$ 
particularly shows three successive transitions; the structural transition 
at $T_{\rm t}$, CO transition at $T_{\rm CO}$ and antiferromagnetic transition at 
$T_{\rm N}$.$^{2,3,5)}$ The magnetic interaction is ferromagnetic above 
$T_{\rm t}$, while below $T_{\rm t}$ it is antiferromagnetic.$^{2,3,5)}$ In 
Y$_{0.5}$Ba$_{0.5}$MnO$_{3}$, on the other hand, there is no evidence or no 
trace of the transitions observed in YBaMn$_{2}$O$_{6}$, except for the 
spin-glass (SG) transition at $T_{\rm G}$ = 30 K. The electrical resistivities of 
$R_{0.5}$Ba$_{0.5}$MnO$_{3}$ ($R$ = Nd, Sm and Y) show semiconductive behaviors, 
as shown in Fig. \ref{f6}. The activation energy $E_{\rm a}$ decreases with decreasing 
the ionic radius of $R^{3 + }$ ion.


The results of magnetic properties of $R_{0.5}$Ba$_{0.5}$MnO$_{3}$ are 
summarized in Fig. \ref{f4} as the phase diagram. Here we compare the phase diagram 
among $R$BaMn$_{2}$O$_{6}$, $R_{0.5}$Ba$_{0.5}$MnO$_{3}$ and 
$R_{0.5}A_{0.5}$MnO$_{3}$ ($A$ = Ca and Sr). The electronic states 
characteristic of perovskite manganites such as AFM(A) and COI(CE) seen in 
$R$BaMn$_{2}$O$_{6}$ and $R_{0.5}A_{0.5}$MnO$_{3}$ ($A$ = Ca and Sr) are absent in 
$R_{0.5}$Ba$_{0.5}$MnO$_{3}$. Instead of these states, magnetic glassy states 
govern the electronic state of $R_{0.5}$Ba$_{0.5}$MnO$_{3}$. The magnetic 
glassy state could be due to a disorder effect that hinders magnetic 
long-range ordering and it could occur as a result of the competition 
between randomly distributed ferromagnetic and antiferromagnetic 
interactions. Since the ionic radius of Ba$^{2 + }$ (= 1.61 {\AA})$^{15)}$ 
is much larger than that of Sr$^{2 + }$ (= 1.44 {\AA})$^{15)}$ and $R^{3 + }$ 
($ \le $1.36 {\AA})$^{15)}$, $R_{0.5}$Ba$_{0.5}$MnO$_{3}$ could include any 
spatial heterogeneity in a nanometer scale, which results in magnetic 
nonhomogeneous states. Only the largest La$^{3 + }$ among $R^{3 + }$s forms a 
homogeneous solid-solution at the $A$-sites with Ba$^{2 + }$ and the magnetic 
long-range ordering of FM is realized in La$_{0.5}$Ba$_{0.5}$MnO$_{3}$.

It has been suggested that the electromagnetic properties of perovskite 
manganites with $A$-site cations randomly distributed depend on not only $f$ but 
also the variance of $A$-cation radius distribution $\sigma ^{2}$ defined as 
$\sigma ^2 = \sum\limits_i {y_i r_i^2 } - r_A^2 $, where $r_{i}$ is the ionic 
radius of each $A$-site cation, $y_{i}$ is the fractional occupancy of the $i$ ion, 
$r_{A}$ is the average ionic radius of $A$-site cations.$^{16)}$ The value of 
$\sigma ^{2}$ indicates the magnitude of potential disorder effect. Here, 
we discuss the ground states of the $A$-site disordered systems 
$R_{0.5}A_{0.5}$MnO$_{3}$ ($A$ = Ca, Sr and Ba) in terms of $\sigma ^{2}$. 
Figure 7 shows the mapping of $R_{0.5}A_{0.5}$MnO$_{3}$ ($A$ = Ca, Sr and Ba) 
on a $\sigma ^{2}$ - $\langle r_{A} \rangle$ diagram. The data of 
$R_{0.5}A_{0.5}$MnO$_{3}$ ($A$ = Sr and Ca) are quoted from the previous 
literatures$^{1,17)}$. The thick lines in Fig. 7 represent possible phase 
boundaries. The magnetic long-range orderings (AFI, AFM and FM) tend to be 
stabilized in the lower $\sigma ^{2}$ region ($\sigma ^{2}$ $<$ 10$^{ - 
2})$; otherwise, the magnetic glassy state is obviously dominant above 
$\sigma ^{2}$ = 10$^{ - 2}$, except FM in La$_{0.5}$Ba$_{0.5}$MnO$_{3}$. 
Thus the difference of the ionic radius between $A$-site cations significantly 
influences magnetic long-range ordering in perovskite manganites. In 
connection with the disordered effect, the lowering of both $T_{\rm C}$ and 
$T_{\rm CO}$ in the critical region is not recognized in the phase diagram of 
$R$BaMn$_{2}$O$_{6}$ in contrast to $R_{0.5}A_{0.5}$MnO$_{3}$ ($A$ = Sr and Ca). 
The absence of such critical behavior in $R$BaMn$_{2}$O$_{6}$ is partly due to 
the $A$-site ordering. In the critical region where FM (AFM) and CO 
interactions compete against each other, they are significantly affected by 
fluctuation of composition, coherent size of crystal and external field 
\textit{etc}, that is the $A$-site randomness, and such fluctuation of interactions 
enhances the criticality. On the other hand, it could be more definite in 
the $A$-site ordered $R$BaMn$_{2}$O$_{6}$ which interaction becomes dominant or which electronic state is stable.

\begin{figure}[tb]
\begin{center}
\includegraphics{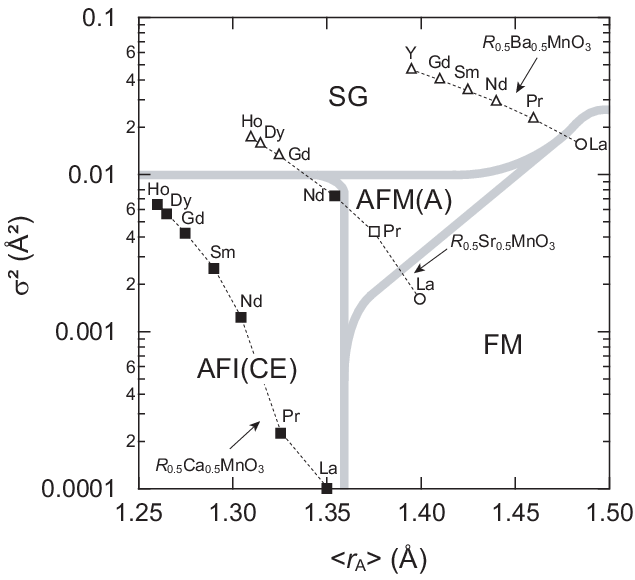}
\end{center}
\caption{A mapping of $R_{0.5}A_{0.5}$MnO$_{3}$ ($A$ = Ca, Sr and Ba) on a $\sigma 
^{2}$ - $\langle r_{A} \rangle$ diagram. (See the text).}
\label{f7}
\end{figure}

\subsection{The $A$-site randomness effect in Pr-compounds}

\begin{figure}[tb]
\begin{center}
\includegraphics{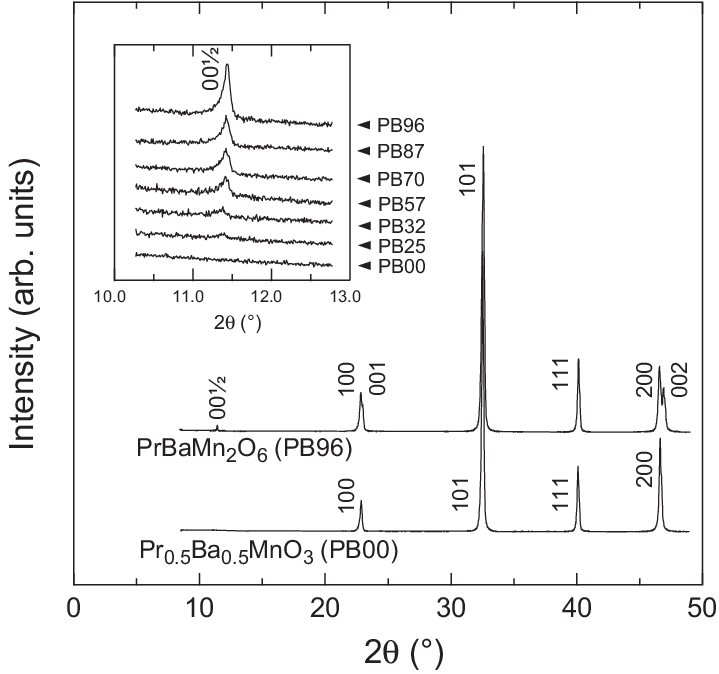}
\end{center}
\caption{X-ray diffraction patterns of the $A$-site ordered PrBaMn$_{2}$O$_{6}$ (PB96) 
and disordered Pr$_{0.5}$Ba$_{0.5}$MnO$_{3}$ (PB00). The inset shows 
$(00\frac{1}{2})_{\rm p}$ reflections of 
Pr-compounds (PB96 $\sim $ PB00) with various degrees of Pr/Ba randomness at 
the $A$-sites.}
\label{f8}
\end{figure}

\begin{figure}
\begin{center}
\includegraphics{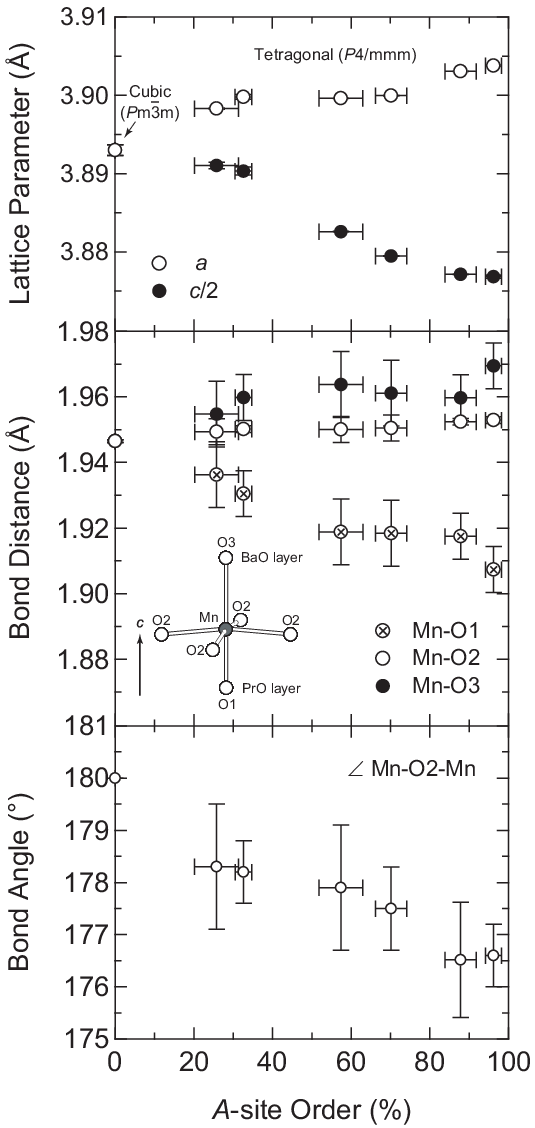}
\end{center}
\caption{Structural parameters at room temperature plotted as a function of the degree of the $A$-site order ({\%}) for Pr-compounds (PB96 $\sim $ PB00).}
\label{f9}
\end{figure}

\begin{figure}
\begin{center}
\includegraphics{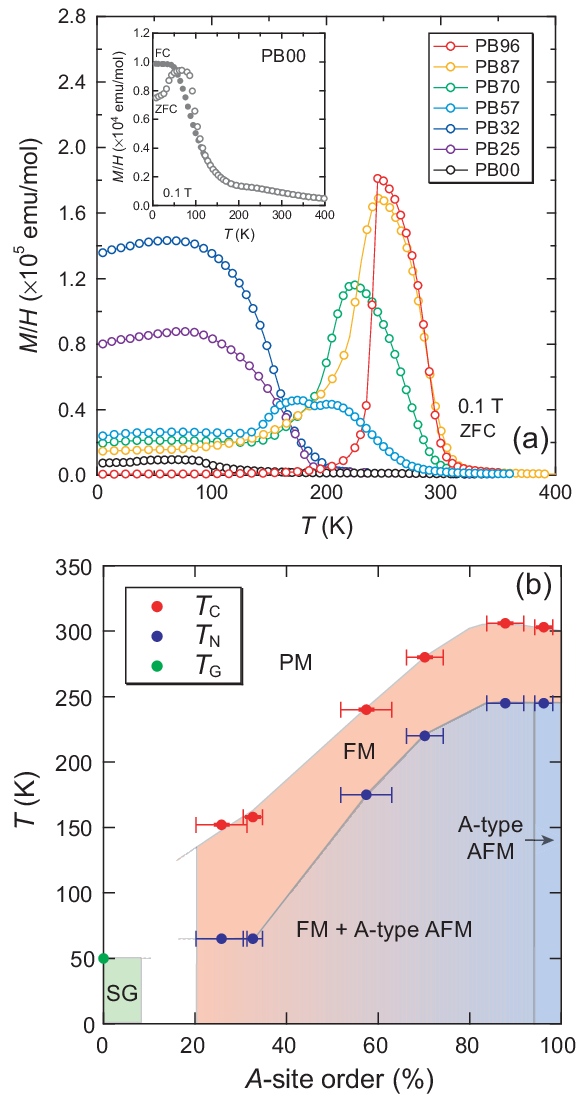}
\end{center}
\caption{(a) Temperature dependence of magnetic susceptibility for Pr-compounds (PB96 
$\sim $ PB00) under 0.1 T. (b) Phase diagram for Pr-compounds as a function 
of the degree of the $A$-site order ({\%}).}
\label{f10}
\end{figure}

As mentioned above, the structure and electromagnetic properties of 
perovskite manganites are significantly affected by the $A$-site disorder. In 
this section, we report the relation between those and the degree of the 
$A$-site disorder in Pr-compounds (PB96 $\sim $ PB00) with various degrees of 
the $A$-site randomness. Figure \ref{f8} shows X-ray diffraction patterns of the 
$A$-site ordered PrBaMn$_{2}$O$_{6}$ (PB96) and disordered 
Pr$_{0.5}$Ba$_{0.5}$MnO$_{3}$ (PB00). The X-ray diffraction pattern can be 
indexed in a simple tetragonal $P$4/\textit{mmm} for PB96 $\sim $ PB25 and a simple cubic 
$Pm\overline 3 m$ for PB00. All of them show no trace of impurity or phase 
separation. The inset shows the $(00 \frac{1}{2} )_{\rm p}$ 
reflections of Pr-compounds (PB96 $\sim $ PB00). The intensity of 
$(00 \frac{1}{2} )_{\rm p}$ reflection decreases 
with increasing the $A$-site randomness and finally it becomes undetectable in 
PB00. Table \ref{t2} and Fig. \ref{f9} show the structure data of Pr-compounds. The lattice 
parameters, $a$ and $c$/2, gradually approach each other with increasing the 
degree of the $A$-site disorder and terminate to the $a$-parameter of cubic PB00. 
In PB96, MnO$_{6}$ octahedra are heavily distorted in a manner as shown in 
Fig. \ref{f9}(b); the apical Mn-O(1) distance is quite short (1.907 {\AA}), while 
the other apical Mn-O(3) distance is long (1.969 {\AA}), as a result of the 
displacement of MnO$_{2}$ plane toward to PrO layer. On the other hand the 
planar Mn-O(2) distance (1.952 {\AA}) is close to the average length of 
Mn$^{3.5 + }$-O bond for conventional perovskite manganites. With increasing 
the $A$-site disorder, MnO$_{6}$ octahedra gradually approach to regular 
octahedra.
\begin{fulltable}[htbp]
\caption{Refined structural parameters Pr-compounds (PB96$\sim$PB00) with various degrees of the $A$-site randomness. (See the text).}
\label{t2}
\begin{fulltabular}{@{\hspace{\tabcolsep}\extracolsep{\fill}}ccllllll} \hline
Sample & &PB96&PB96&PB87&PB70&PB57\\
\hline
Data Resource & & X-ray & Neutron & X-ray & X-ray & X-ray\\
Space Group &  &$P$4/$mmm$&$P$4/$mmm$&$P$4/$mmm$&$P$4/$mmm$&$P$4/$mmm$\\
$a$ (\AA) & & 3.9038(1) & 3.9029(1) & 3.90309(5)& 3.89998(5)& 3.89961(5)\\
$c$ (\AA) & & 7.7537(4) & 7.7547(5) & 7.75432(8) & 7.7589(1)& 7.7651(1)\\
$g$ & & 0.98(1) & 0.98(1) & 0.93(2) & 0.85(2) & 0.78(3)\\
Pr/Ba at [ ]$_{\rm P}$ & 1$a$ & $B$$_{\rm iso}$=0.8(1)/0.5 & $B$$_{\rm iso}$=0.8(1)/0.6 & $B$$_{\rm iso}$=0.9(1)/0.6 & $B$$_{\rm iso}$=0.9(1)/0.5 & $B$$_{\rm iso}$=0.9(1)/0.6\\
Pr/Ba at [ ]$_{\rm B}$ & 1$b$ & $B$$_{\rm iso}$=0.5(1)/0.8 & $B$$_{\rm iso}$=0.6(1)/0.8 & $B$$_{\rm iso}$=0.6(1)/0.9 & $B$$_{\rm iso}$=0.5(1)/0.9 & $B$$_{\rm iso}$=0.6(1)/0.9\\
Mn & 2$h$ & $z$=0.2460(7) & $z$=0.2463(6) & $z$=0.2472(9) & $z$=0.247(1) & $z$=0.247(2)\\
 & & $B$$_{\rm iso}$=0.2(1) & $B$$_{\rm iso}$=0.25(8) & $B$$_{\rm iso}$=0.2(1) & $B$$_{\rm iso}$=0.3(1) & $B$$_{\rm iso}$=0.2(1)\\
O1 & 1$c$ & $B$$_{\rm iso}$=1.0(1) & $B$$_{\rm iso}$=0.9(1) & $B$$_{\rm iso}$=1.0(1) & $B$$_{\rm iso}$=1.1(1) & $B$$_{\rm iso}$=1.1(1)\\
O2 & 4$i$ & $z$=0.2386(4) & $z$=0.2385(3) & $z$=0.239(1) & $z$=0.241(1) & $z$=0.242(2)\\
 & & $B$$_{\rm iso}$=1.0(1) & $B$$_{\rm iso}$=1.0(1) & $B$$_{\rm iso}$=1.0(1) & $B$$_{\rm iso}$=1.1(1) & $B$$_{\rm iso}$=1.1(1)\\
O3 & 1$d$ & $B$$_{\rm iso}$=1.0(1) & $B$$_{\rm iso}$=1.0(1) & $B$$_{\rm iso}$=1.0(1) & $B$$_{\rm iso}$=1.1(1) & $B$$_{\rm iso}$=1.1(1)\\
$R$$_{\rm wp}$ (\%) & & 10.29 & 8.95 & 12.02 &13.68 &12.40\\
$R$$_{\rm e}$ (\%) & & 7.75 & 6.79 & 8.64 & 9.02 & 8.71\\
$A$-site Order (\%) & & 96(2) & 96(2) & 87(4) & 70(4) & 57(6)\\
\end{fulltabular}
\begin{fulltabular}{@{\hspace{\tabcolsep}\extracolsep{\fill}}cclllll} \hline
Sample & &PB57&PB32&PB32&PB25&PB00\\
\hline
Data Resource & & Neutron & X-ray & Neutron & X-ray & X-ray\\
S.G. &  &$P$4/$mmm$&$P$4/$mmm$&$P$4/$mmm$&$P$4/$mmm$&$Pm$$\bar3$$m$\\
$a$ (\AA) & & 3.8997(1)& 3.8998(3) & 3.8996(2) & 3.89829(9) & 3.8930(7) \\
$c$ (\AA) & & 7.7658(3) & 7.7806(8) & 7.7806(5)& 7.7820(4) & \\
$g$ & &0.79(2) & 0.66(1) & 0.64(1) & 0.63(3) & 0.5\\
Pr/Ba at [ ]$_{\rm P}$ & 1$a$ & $B$$_{\rm iso}$=0.8(1)/0.4 & $B$$_{\rm iso}$=0.9(1)/0.6 & $B$$_{\rm iso}$=0.8(1)/0.5 & $B$$_{\rm iso}$=0.9(1)/0.6 & Pr/Ba 1$a$\\
Pr/Ba at [ ]$_{\rm B}$ & 1$b$ & $B$$_{\rm iso}$=0.4(1)/0.8 & $B$$_{\rm iso}$=0.6(1)/0.9 & $B$$_{\rm iso}$=0.5(1)/0.8 & $B$$_{\rm iso}$=0.6(1)/0.9 & $B$$_{\rm iso}$=0.8(1)/0.6(1)\\
Mn & 2$h$ & $z$=0.2474(7) & $z$=0.248(2) & $z$=0.2478(8) & $z$=0.249(1) & Mn 1$b$\\
 & & $B$$_{\rm iso}$=0.29(8) & $B$$_{\rm iso}$=0.2(1) & $B$$_{\rm iso}$=0.2(1) & $B$$_{\rm iso}$=0.3(1) & $B$$_{\rm iso}$=0.31(7)\\
O1 & 1$c$ & $B$$_{\rm iso}$=1.2(1) & $B$$_{\rm iso}$=1.1(1) & $B$$_{\rm iso}$=1.1(1) & $B$$_{\rm iso}$=1.0(1) & \\
O2 & 4$i$ & $z$=0.2422(9) & $z$=0.2440(7) & $z$=0.2439(7) & $z$=0.245(1) & O 3$c$\\
 & & $B$$_{\rm iso}$=1.2(1) & $B$$_{\rm iso}$=1.1(1) & $B$$_{\rm iso}$=1.1(1) & $B$$_{\rm iso}$=1.0(1) & $B$$_{\rm iso}$=1.2(1)\\
O3 & 1$d$ & $B$$_{\rm iso}$=1.2(1) & $B$$_{\rm iso}$=1.1(1) & $B$$_{\rm iso}$=1.1(1) & $B$$_{\rm iso}$=1.0(1) & \\
$R$$_{\rm wp}$ (\%) & & 10.09 & 11.32 & 9.65 & 12.05 &10.69\\
$R$$_{\rm e}$ (\%) & & 6.99 & 8.59 & 6.41 & 9.27 & 8.65\\
$A$-site Order (\%) & & 58(4) & 32(2) & 29(2) & 25(6) & 0\\
\hline
\end{fulltabular}
\footnotetext{a}{\small The degree of $A$-site order ($S$) is defined by $(2g-1)\cdot 100 (\%)$, where $g$ is the refined occupancy factor represented as [Pr$_g$Ba$_{1-g}$]$_{\rm P}$[Pr$_{1-g}$Ba$_g$]$_{\rm B}$Mn$_{2}$O$_{6}$; [ ]$_{\rm P}$ and [ ]$_{\rm B}$ show Pr-sites and Ba-sites in PrBaMn$_{2}$O$_{6}$, respectively.}\\
\end{fulltable}
The magnetic susceptibilities ($M$/$H)$ of Pr-compounds measured under 0.1 T are 
shown in Fig. \ref{f10}(a). PB96 shows FM transition at $T_{\rm C}$ = 303 K, followed by 
AFM(A) transition at $T_{\rm N}$ = 245 K. With increasing the $A$-site randomness, 
both $T_{\rm C}$ and $T_{\rm N}$ slightly decrease and AFM(A) transitions become 
broad. Furthermore some amount of FM state coexists with AFM(A) state below 
$T_{\rm N}$ in PB87 and PB70, which is evidenced by a considerable amount of 
temperature independent $M/H$ below $T_{\rm N}$. On the other hand, PB32 and PB25 
with a considerable $A$-site disorder exhibit clear FM transitions at $T_{\rm C}$ = 
158 K and 152 K, respectively. Since the saturated values of $M/H$ of PB32 and 
PB25 are lower than that expected from full moment, any short-range magnetic 
ordered phase and/or AFM(A) phase coexist with FM phase. The rather low 
$T_{\rm C}$s for the second group (PB32 and PB25) compared with $T_{\rm C}$s for the 
first group (PB96-PB70) suggests two types of FM phase in Pr-compounds. This 
might be reflected in the $M/H-T $curve with two peaks around 200 K and 180 K for 
the intermediate compound PB57, namely PB57 includes two FM phases and each 
FM phase transforms to AFM(A) phase at independent temperatures (200 K and 
180 K), showing peaks in $M/H$-$T$ curve. Finally, the perfect disordered PB00 has 
a much smaller $M/H$ than that of other compounds and shows a spin glass like 
transition at 50 K, as shown in the inset of Fig. \ref{f10}(a). A small amount of 
AFM(A) phase was confirmed at low temperature by a neutron diffraction 
measurements, but FM long-range ordered phase was not observed. 

The obtained results of Pr-compounds are summarized in Fig. \ref{f10}(b). The 
$A$-site order stabilizes AFM(A) state associated with a $d_{x^2 - y^2} $ 
orbital order (layer type), because the layer type order of $R$/Ba and 
consequently the distorted MnO$_{6}$ octahedra introduce 2-dimensionality in 
the crystal structure. The increase of the $A$-site disorder makes AFM(A) state 
unstable because of the decrease of structural anisotropy 
(2-dimensionality). On the other hand, it is advantageous to FM state 
generated by an isotropic double exchange interaction and it leads to FM 
state for PB32 and PB25. However, the effect of $\sigma ^{2}$ is 
simultaneously enhanced by the $A$-site disorder and finally results in 
magnetic glassy state in the disordered form PB00. In conclusion the 
$A$-site randomness in Ba-based manganites clearly suppresses not only FM 
transition but also AFM(A) transition and leads to magnetic glassy state. We 
observed a similar randomness effect on Nd-compounds. Previously Akahoshi 
\textit{et al.}$^{11)}$ reported somewhat different result of FM states as the ground state 
for Pr$_{0.5}$Ba$_{0.5}$MnO$_{3}$ and Nd$_{0.5}$Ba$_{0.5}$MnO$_{3}$. The 
present experiments suggest insufficient disorder in their samples.

\begin{figure}
\begin{center}
\includegraphics{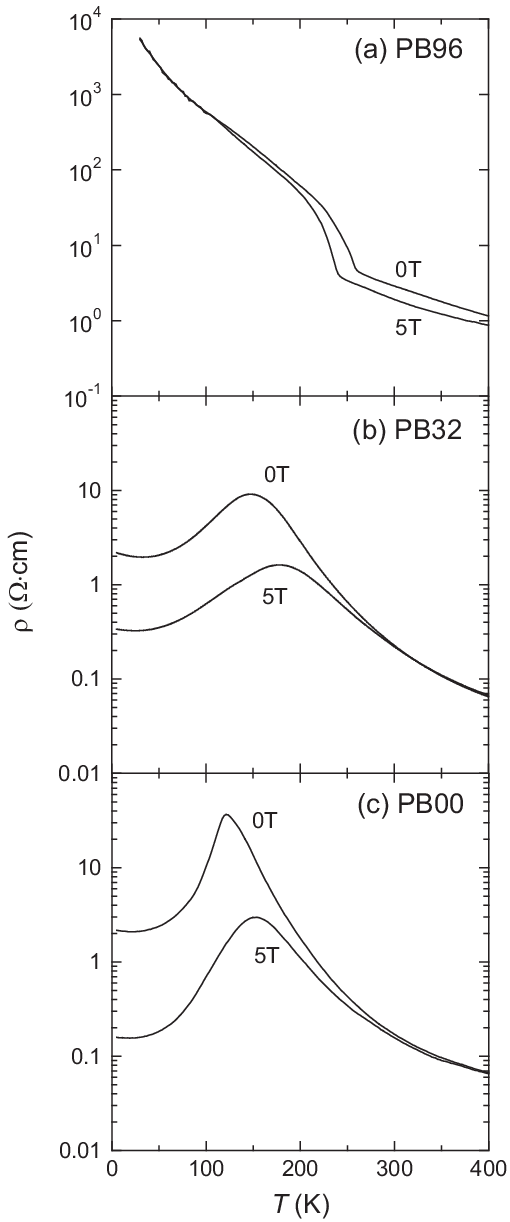}
\end{center}
\caption{Temperature dependence of electrical resistivity at 0 and 5 T for (a) PB96, 
(b) PB32 and (c) PB00.}
\label{f11}
\end{figure}

\begin{figure}
\begin{center}
\includegraphics{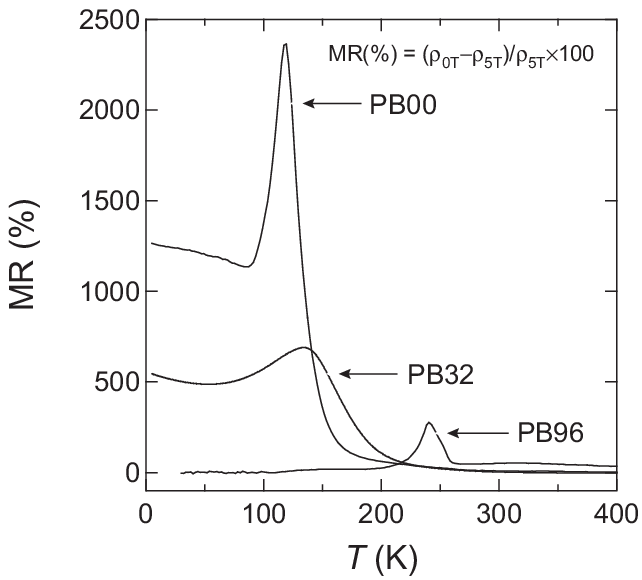}
\end{center}
\caption{Magnetoresitance (MR) vs. temperature plots for PB96, PB32 and PB00. MR 
effect was measured between 0 and 5 T.}
\label{f12}
\end{figure}

Figure \ref{f11} shows temperature variation of electrical resistivity $\rho $ in (a) 
PB96, (b) PB32 and (c) PB00 at 0 and 5 T. For PB96 in which AFM(A) state is 
stable, magnetic field dependence of resistivity is little, although AFM(A) 
transition temperature is obviously suppressed by 18 K at 5 T. On the other 
hand, magnetoresistance (MR) effect is observed below $T_{\rm C}$ ( = 158 K) for 
ferromagnetic PB32 and below 120 K even for PB00 which has no long-range 
ferromagnetic order. The MR effects of these compounds at 5 T are summarized 
in Fig. \ref{f12}, where MR({\%}) is given by 
$\rm {MR}(\%) = \{[\rho(0)-\rho(H)]/\rho(0)\} \times 100\%$ with $\rho (H)$ in 5 T and $\rho (0)$ in zero magnetic 
field. With increasing the $A$-site randomness, MR effect increases and the 
maximum MR effect reaches to 2360 {\%} in PB00, although the temperature 
($T_{\rm MR})$ at the maximum MR effect decreases. It is obvious that the 
$A$-site randomness increases MR effect in Pr-compounds. This is the first 
observation of the efficient MR effect caused by the $A$-site disorder in a 
series of compounds with a fixed composition and various degrees of the 
$A$-site randomness. We also found very similar behaviors in Nd-compounds.

\begin{figure}
\begin{center}
\includegraphics{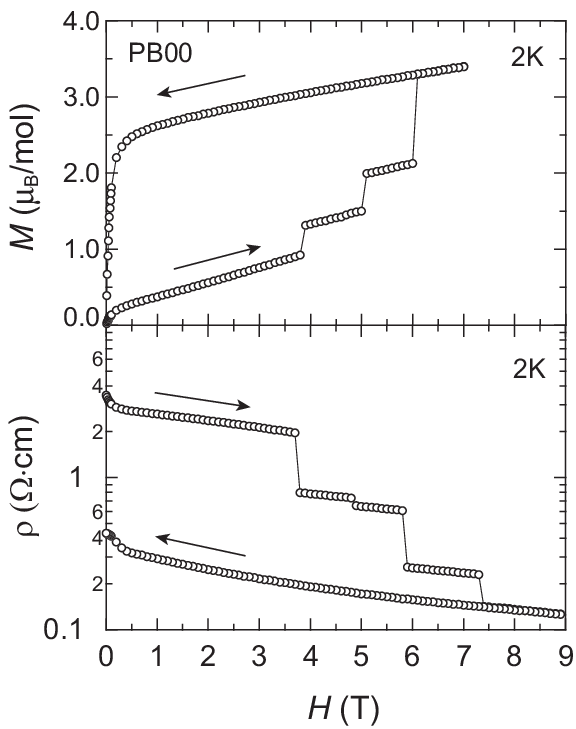}
\end{center}
\caption{Magnetic field dependences of resistivity and magnetization at 2 K for PB00.}
\label{f13}
\end{figure}
The decrease of resistivity below 150 K in PB32 is due to the development of 
FM state and the observed MR effect could be associated with the conversion 
of coexisted phases (AFM(A) phase or magnetic glassy phase) to FM phase by 
an external magnetic field. On the other hand PB00 shows no long range 
magnetic ordering nevertheless it shows similar behaviors of resistivity and 
MR effect. Such behaviors would be due to the development of short range 
magnetic ordering and the conversion of magnetic glassy phase to FM phase by 
an external magnetic field. At 2 K in PB00, a peculiar behavior, as shown in 
Fig. \ref{f13}, has been observed. The resistivity decreases stepwise as the 
magnetic field increases, while the magnetization increases stepwise with 
the close relation to the resistivity behaviors. These behaviors are not 
reversible in the magnetic field. The stepwise behaviors in the 
magnetization and resistivity were observed up to 4.9 K but they vanished 
dramatically at 5.0 K. Similar behaviors were previously reported in 
Pr$_{0.5}$Ca$_{0.5}$MnO$_{3}$ doped with a few percent of other cations such 
as Sc, Ga or Co on the Mn site and were explained by an impurity 
induced-disorder, with the coexistence of several short-range AFI(CE) phases 
and small FM regions.$^{18,19)}$ Our system has neither FM-to-AFI(CE) 
transition nor dopant in contrast to Pr$_{0.5}$Ca$_{0.5}$MnO$_{3}$. A model 
based on ordinal two-phase mixture cannot explain the behavior. For 
instance, AFI(CE) phase in the coexistence with FM phase is continuously 
converted to FM phase as observed in LaBaMn$_{2}$O$_{6}$.$^{6)}$ We have no 
explanation for such multi-step magnetization and resistivity change at 
present. However we would like to emphasize a close relation between the 
observed behavior and any spatial heterogeneity in a nanometer scale. 
Detailed study is now in progress.

\section{Summary}

To summarize, we have investigated the structures and electromagnetic 
properties of the $A$-site disordered Ba-based manganite 
$R_{0.5}$Ba$_{0.5}$MnO$_{3}$ ($R$ = Y and rare earth elements) and compared 
$R_{0.5}$Ba$_{0.5}$MnO$_{3}$ with not only the $A$-site ordered manganite 
$R$BaMn$_{2}$O$_{6}$ but also ordinary disordered manganites 
$R_{0.5}A_{0.5}$MnO$_{3}$ ($A$ = Ca and Sr). The disordered form 
$R_{0.5}$Ba$_{0.5}$MnO$_{3}$ has a primitive cubic perovskite cell with no 
tilt of MnO$_{6}$ octahedra. The electronic states characteristic of 
perovskite manganites are absent in $R_{0.5}$Ba$_{0.5}$MnO$_{3}$ and magnetic 
glassy states govern the electronic state of $R_{0.5}$Ba$_{0.5}$MnO$_{3}$. 
The magnetic glassy states could be due to the disorder effect that hinders 
the long-range magnetic ordering and could occur as a result of the 
competition between randomly distributed ferromagnetic and antiferromagnetic 
interactions. The $A$-site randomness effect has been investigated in 
Pr-compounds with various degrees of Pr/Ba randomness at the $A$-sites. The 
$A$-site randomness suppresses both ferromagnetic and $A$-type antiferromagnetic 
transitions in PrBaMn$_{2}$O$_{6}$. On the other hand, magnetoresistance 
effect becomes remarkable with increase of the $A$-site disorder. As remarkable 
phenomena, multi-step magnetization and resistivity changes have been 
observed in Pr$_{0.5}$Ba$_{0.5}$MnO$_{3}$. Since the ionic radius of Ba$^{2 
+ }$ is much larger than that of Sr$^{2 + }$ and also $R^{3 + }$, 
$R_{0.5}$Ba$_{0.5}$MnO$_{3}$ could include any spatial heterogeneity in a 
nanometer scale, which could be closely related to the multi-step 
magnetization and resistivity changes observed.

\section*{Acknowledgements}
The authors thank T. Yamauchi, M. Isobe, Y. Matsushita, H. Kageyama and K. 
Ueda for valuable discussion. This work is partly supported by Grants-in-Aid 
for Scientific Research (No. 407 and No. 758) and for Creative Scientific 
Research (No. 13NP0201) from the Ministry of Education, Culture, Sports, 
Science, and Technology.

\end{document}